\documentstyle[aps,preprint]{revtex}

\begin{document}
\title{Antiferromagnetic Correlations versus Superfluid Density in La$_{2-x}$Sr$_x$CuO$_4$}

\author{C. Panagopoulos$^a$, B.D. Rainford$^b$, J.R. Cooper$^a$, C.A. Scott$^c$}

\address{$^a$ IRC in Superconductivity and Department of Physics, University of Cambridge, Cambridge CB3 0HE, United Kingdom}

\address{$^b$ Department of Physics and Astronomy, University of Southampton, Southampton  S017 1BJ, United Kingdom}

\address{$^c$ Rutherford Appleton Laboratory, ISIS Facility, Didcot, Oxon OX11 0QX, United Kingdom}

\maketitle

\begin{abstract}
We have performed muon spin relaxation and low field $ac$-susceptibility measurements in a series of high quality samples of La$_{2-x}$Sr$_x$CuO$_4$ ($x=0.08-0.24$) as a function of temperature. Superconductivity is found to coexist with low temperature spin glass order up to the optimally doped region where the normal state pseudogap also closes. The systematic depletion of the superfluid density with the enhancement of aniferromagnetic correlations with underdoping indicates a $competition$ between antiferromagnetic correlations and superconductivity.

\end{abstract}

Establishing and understanding the phase diagram of the high-$T_c$ superconductors (HTS) versus temperature and doping has been one of the major challenges in modern solid state physics. The parent compound La$_2$CuO$_4$ of the first HTS family to be discovered, La$_{2-x}$Sr$_x$CuO$_4$, is an insulator exhibiting long range antiferromagnetic (AF) order, which is eventually destroyed as carriers are doped into the CuO$_2$ planes. After passing through a spin glass (SG) state superconductivity emerges near 0.06 holes/planar-Cu-atom and follows an approximately parabolic doping dependence until it disappears at $x\simeq$0.30. Spectroscopic evidence however, indicates that AF correlations do not seize to exist with the emergence of superconductivity, but instead a short range ordered AF state persists in the superconducting state \cite{KY,CN}. This observation raises fundamental questions as to how far into the superconducting regime of the phase diagram these AF correlations persist and how, if at all, these affect the superconductivity.

In this short paper we present briefly new experimental findings in which the signatures of antiferromagnetic correlations is observed to persist up to $x\simeq$0.17. We find a close correlation between the doping dependence of the AF and the absolute value of the superfluid density, $\rho^s(0)$. The latter is almost constant for approximately $x>0.17$ and drops with the enhancement of the AF fluctuations, undergoing an increased depletion in the vicinity of the stripe phase region ($x=0.125$). These results suggest that the AF order parameter competes with the superconducting counterpart in more than $50\%$ of the superconducting region of the phase diagram.

The samples studied were single-phase polycrystalline La$_{2-x}$Sr$_x$CuO$_4$ (LSCO)  ($x=$ 0.08, 0.10, 0.125, 0.15, 0.17, 0.20, 0.22, 0.24) prepared using solid-state reaction procedures. No other phases were detected by powder x-ray diffraction and the phase purity is thought to be better than $1\%$. High field magnetic susceptibility measurements showed no signatures of excess paramagnetic centres and the measured values of $T_c$ and lattice parameters are also in very good agreement with published work \cite{Rad}. The samples have been extensively characterised by several transport, magnetic and spectroscopic techniques, all indicating their high quality. Zero-field (ZF) and transverse-field (TF) $\mu$SR experiments were performed at the pulsed muon source, ISIS Facility, Rutherford Appleton Laboratory. The samples were mounted on a silver plate either on the cold stage of a dilution refrigerator or in a variable temperature helium cryostat, enabling spectra to be collected over the temperature range $40mK$ to $50K$. The in-plane magnetic penetration depth, $\lambda _{ab}$, ($\lambda _{ab}^{-2}$ $\sim$ $\rho^s$) was determined both by a low-field $ac$-susceptibility technique (typically at $1G$ and $333Hz$) on grain-aligned powders and from the analysis of TF-cooled $\mu$SR measurements. The latter were carried out on unaligned powders in a field of $400G$. Details for deriving $\lambda _{ab}$ in HTS from the measured low-field $ac$-susceptibility and TF-$\mu$SR spectra can be found elsewhere \cite{CP1,CP2}.

In Fig.1 we present the time evolution of the ZF muon asymmetry for $x=0.08$ ($T_c=21K$) as a function of temperature. In all samples the high temperature form of the depolarisation is Gaussian, consistent with dipolar interactions between the muons and their near neighbour nuclear moments. This was verified by applying a $50G$ longitudinal field, which completely suppressed the depolarisation. As the temperature is lowered the onset of dynamical relaxation processes becomes apparent in the change in the shape of the depolarisation function. The samples with $x=0.08, 0.10$ and $0.125$ follow the same pattern, which is indicative of the onset of spin glass ordering at low temperature. For simplicity, we have chosen to parametrise the form of the depolarisation function as a stretched exponential, $G_z(t)$ = $A_1$ $exp[-(\lambda t )^\beta]$ + $A_2$. The constant term $A_2$ accounts for a small time independent background arising from muons stopping in the silver backing plate. At high temperatures we find that the value of $\beta$ $\approx$ 2, but for samples with $x<0.15$ it decreases smoothly to a value approaching $0.5$ at low temperatures. This "root exponential" behaviour is widely found in the temperature regime just above the glass temperature in spin glasses \cite{BDR}. The temperature dependence of $\beta$ for the present samples is shown in Fig. 2. We have used the temperature at which the value of $\beta$ drops below 2 as the onset temperature for AF correlations $T_{sf}$, and the temperature where $\beta$ $\approx$ 0.5 as a measure of $T_{sg}$ the spin glass freezing temperature. We find that for the $x<0.08$ sample for example, the relaxation rate parameter $\lambda^ \beta$ is peaked close to $T_{sg}$ (Fig. 3). At temperatures below $T_{sg}$ the form of the depolarisation function changes: there is an initial rapid decay of $G_z(t)$, followed by a slowly damped tail (see left hand panel in Fig. 1). This behaviour is very characteristic of the behaviour of spin glasses below $T_{sg}$, and has been attributed by Uemura $et$ $al$,  \cite{YU} to the effects of the static distribution of field, combined with dynamical processes in the frozen spin glass state.

The present data show that SG freezing persists up to and beyond $x=0.125$. Indeed the onset of AF correlations for $x=0.125$ occurs at a higher temperatures than for $x=0.10$. This is probably associated with the formation of stripe domains in this range of concentration \cite{KY}. For $x=0.15$ and $0.17$ the trends of $\beta(T)$ suggest a very small value of spin glass temperature $T_{sg}$ ($<45mK$) with fluctuations setting just below 8K and 2K, respectively. This suggests that for higher dopings the AF correlations are absent or at least beyond experimental range.

In Fig. 4 we summarise the essence of this work by comparing the doping dependence of $T_{sf}$ and $T_{sg}$, (including data for $T_{sg}$ from ref. \cite{CN}), with that of $\lambda _{ab}(0)^{-2}$ $\sim$ $\rho^s(0)$. We note that our vales of $T_{sg}$ for $x=0.08$ and $0.10$ are in excellent agreement with those reported in ref. \cite{CN}. Figure 4 indicates that although the freezing of spins occurs at very low temperatures, $T_{sg}$ $\ll$ $T_c$, magnetic fluctuations are apparent at significantly higher temperatures ($e.g.$, $T_{sf}\approx0.5T_c$ for $x=0.10$ and $T_{sf}\approx0.2T_c$ for $x=0.15$). Therefore, a large fraction of the supeconducting region of the phase diagram coexists with AF picking up near $x=1/8$ and eventually disappearing in the lightly overdoped region where the normal state (or pseudo) gap, $\Delta_N$, is known to close \cite{BV}, suggesting a connection between $\Delta_N$ and antiferromagnetic correlations.

As shown in Fig. 4, the superfluid density is doping independent in the region where $T_{sg}$ = $T_{sf}$ $=0$ and is gradually reduced with the increased evolution of AF correlations ($ie$., for $x<0.20$), undergoing a local dip in the $1/8$ region where AF is enhanced possibly due to the stripe phase. We would like to note the striking parallel changes in $\rho^s(0)$ with $T_{sg}$, $T_{sf}$ emphasising the intimate connection of $\rho^s(0)$ with the AF background rather than simply $T_c$.

The present results suggest that AF coexists with superconductivity in the underdoped and slightly overdoped samples and becomes undetectable in the heavily overdoped regime where the pseudogap also disappears. The systematic depletion of the superfluid density with increasing AF correlations (Fig. 4) indicates that the latter $compete$ with superconductivity and $\rho^s(0)$ is not simply a function of $T_c$

We are grateful to Dr A.D. Taylor of the ISIS Facility, Rutherford Appleton Laboratory for the allocation of muon beam time. C.P. thanks Tao Xiang for useful discussions and Trinity College, Cambridge for financial support.

FIGURE CAPTIONS

Figure 1. Typical zero-field $\mu$SR spectra of La$_{2-x}$Sr$_x$CuO$_4$  for $x=0.08$ measured at $1.3, 5$ and $9K$. The solid lines are fits of the stretched exponential,  $G_z(t)$ = $A_1$ $exp[-(\lambda t )^\beta]$ + $A_2$ (see text for details).

Figure 2. Temperature dependence of the exponent $\beta$ of La$_{2-x}$Sr$_x$CuO$_4$  for $x=0.08, 0.10, 0.125, 0.15, 0.17$.

Figure. 3.  Temperature dependence of the relaxation rate parameter $\lambda^\beta$ of La$_{2-x}$Sr$_x$CuO$_4$  for $x=0.08, 0.10, 0.125, 0.15, 0.17$.

Figure 4. Phase diagram of La$_{2-x}$Sr$_x$CuO$_4$ showing the doping dependence of $T_c$ (closed lower triangles), $T_{sg}$ (closed circles), $T_{sf}$ (closed squares), and $\lambda _{ab}(0)^{-2}$ $\sim$ $\rho^s(0)$ (closed upper triangles). Open cirlces are data for $T_{sg}$ taken from ref. [2]. A schematic variation of the Nell temperature $T_N$ is also shown as a broken line.


\begin{references}
\bibitem{KY}  K. Yamada $et$ $al.$, Phys. Rev. B {\bf 57}, 6165 (1998).


\bibitem{CN}  Ch. Niedermayer $et$ $al.$, Phys. Rev. Lett. {\bf 80}, 3843 (1998).


\bibitem{Rad}  P.G. Radaelli $et$ $al.$, Phys. Rev. B {\bf49}, 4163 (1994). 


\bibitem{CP1}  C. Panagopoulos $et$ $al.$, Phys. Rev. Lett. {\bf79}, 2320 (1997).


\bibitem{CP2}  C. Panagopoulos$et$ $al.$, Phys. Rev. B {\bf60}, 14617 (1999). 


\bibitem{BDR}  R. Cywinski and B.D. Rainford, Hyperfine Interactions {\bf85}, 215 (1994).


\bibitem{YU}  Y.J. Uemura $et$ $al.$, Phys. Rev. B {\bf31}, 546 (1985).


\bibitem{BV}  B. Batlogg and C. Varma, Physics World {\bf13}, 33 (2000).

\end{references}
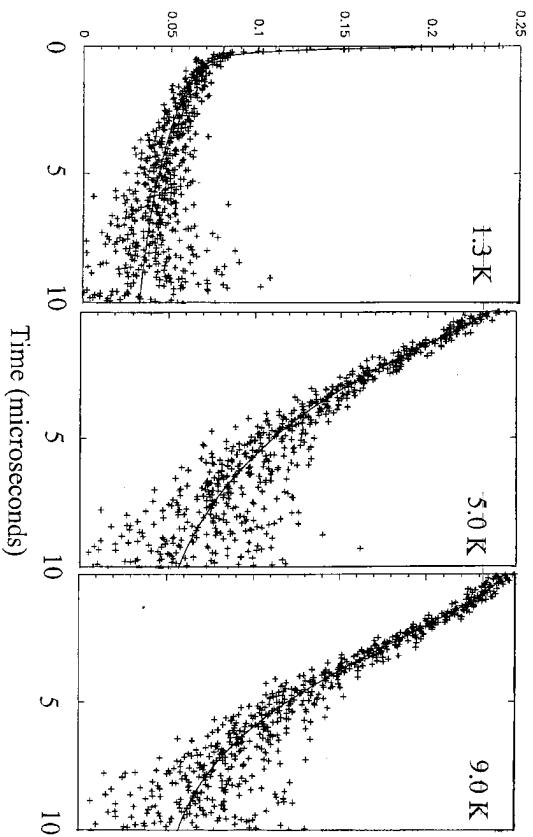
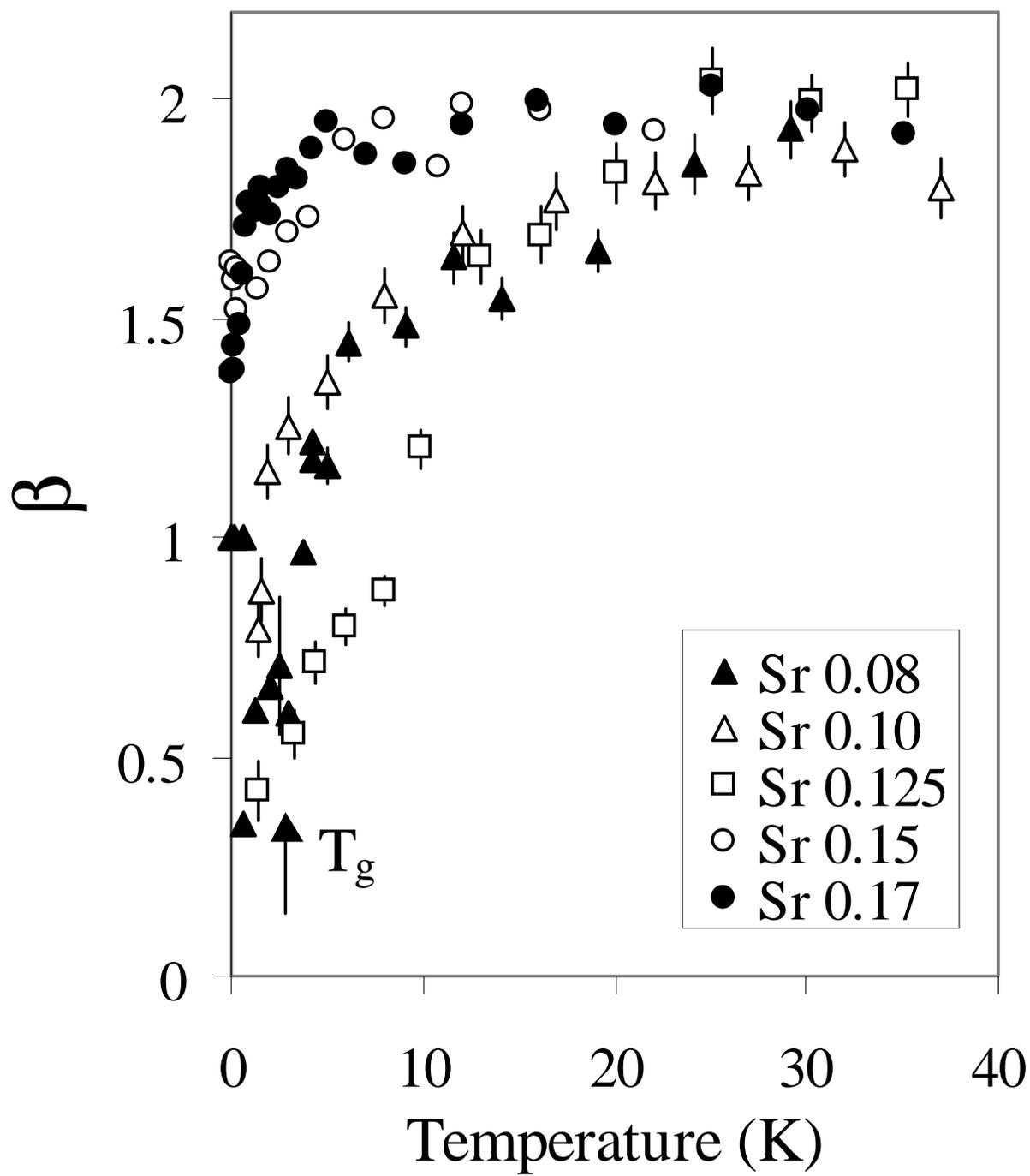
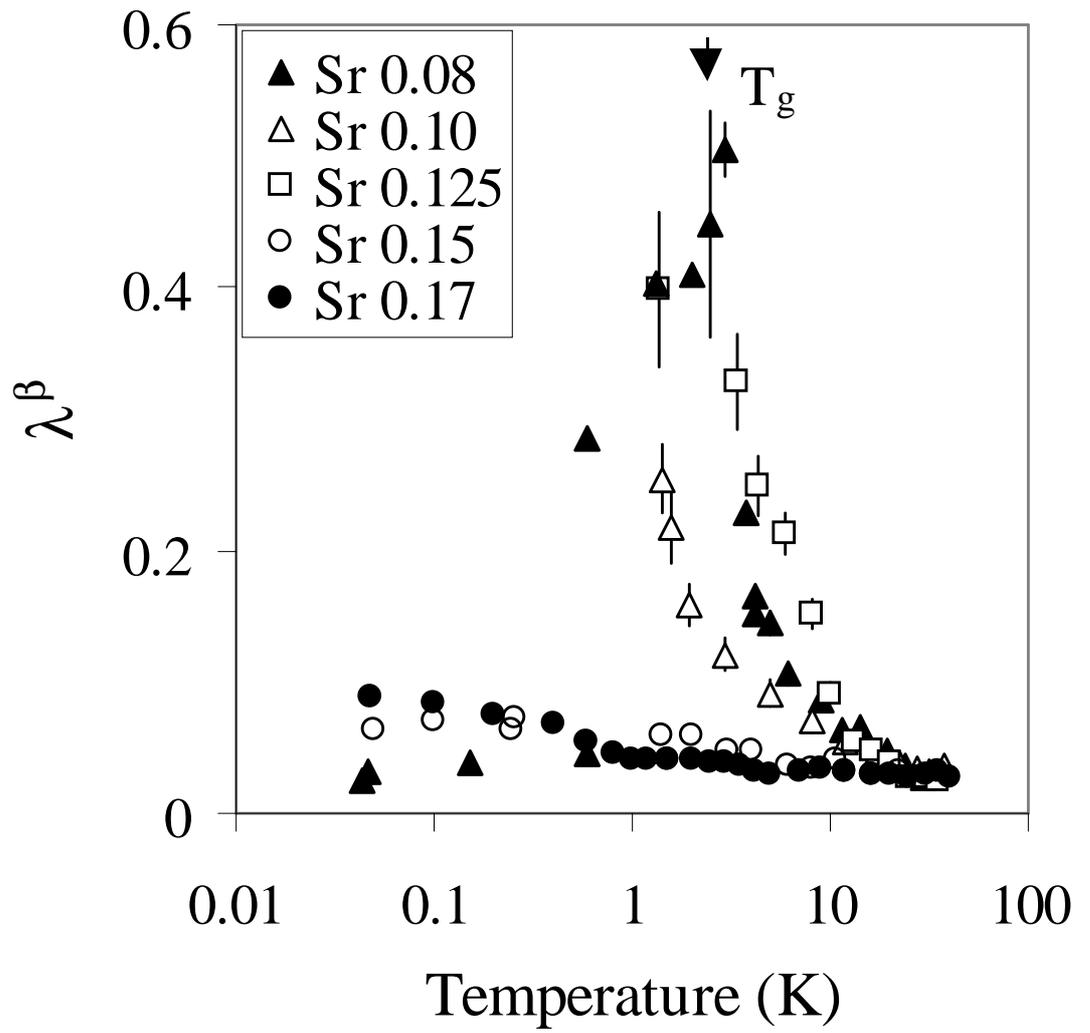
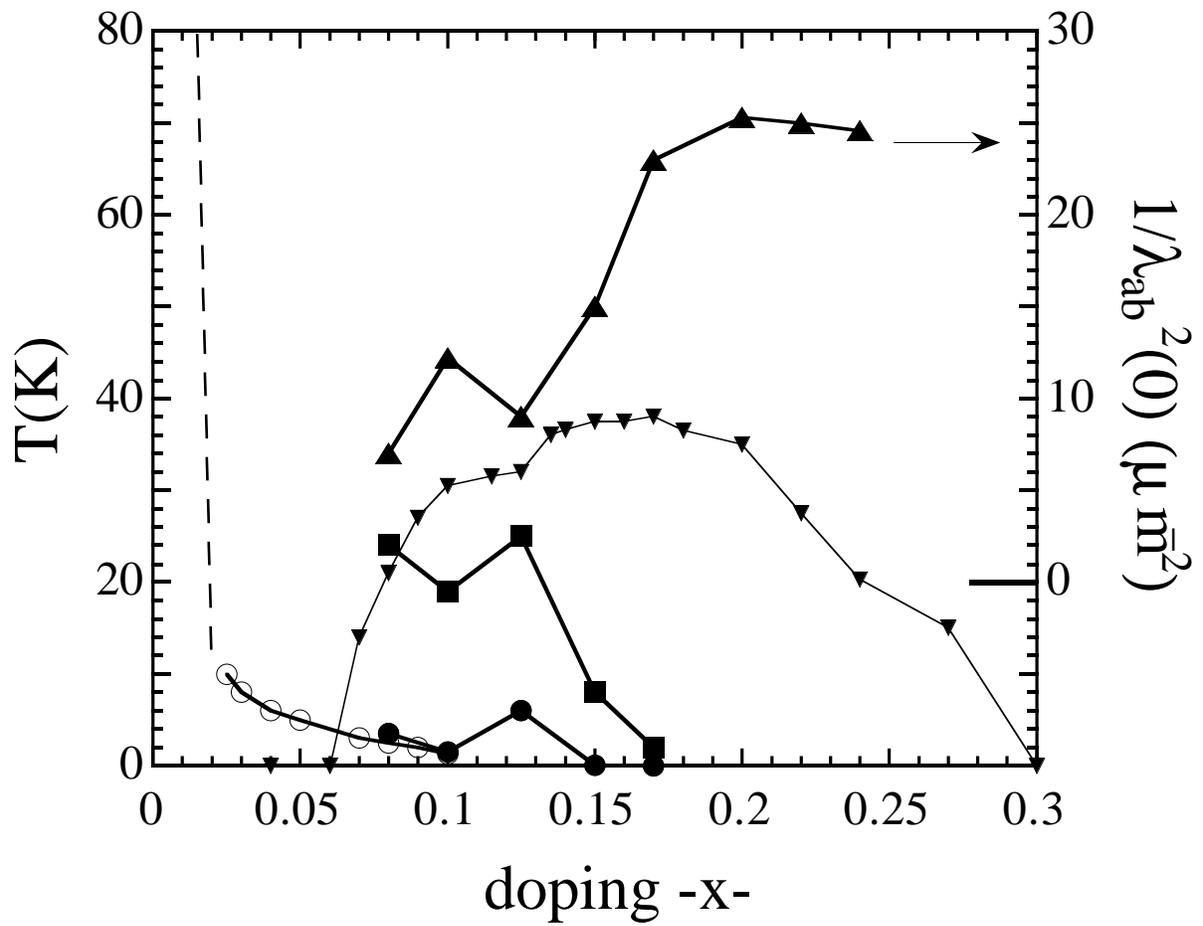
\end{document}